\documentstyle{article}

\newcounter{eqnletter}[equation]
\catcode`\@=11
\@addtoreset{equation}{section}
\catcode`\@=12

\def\parno{\par\noindent}

\begin{document}
\begin{center}
\hfill UPRF-2001-18\\[1cm]
\end{center}
{\centerline {\LARGE {\bf Yang-Mills Integrals}}}
\vskip 1 cm
\centerline  {Giovanni M.Cicuta$\,^a$, Luca Molinari$\,^b$, Graziano Vernizzi$\,^c$}
\vskip .5 cm
{\small \centerline  {$^a$ Dip. di Fisica, Univ. di Parma, Parco Area delle Scienze 7A, 43100 Parma, Italy }
\centerline  {and INFN, Sez.di Milano, Gruppo di Parma}
\centerline  {{\it E-mail:} {\tt cicuta@fis.unipr.it}}
\centerline  {$^b$ Dip. di Fisica, Univ. di Milano, Via Celoria 16, 20133 Milano, Italy}
\centerline  {and INFN, Sez. di Milano}
\centerline  {{\it E-mail:} {\tt Luca.Molinari@mi.infn.it}}
\centerline {$^c$ Dept. of Theoretical Physics, Oxford Univ.}
\centerline  {1 Keble Road, Oxford, OX1 3NP, United Kingdom}
 \centerline {\it E-mail: {\tt vernizzi@thphys.ox.ac.uk}}
\vskip .7 cm
{\centerline{\bf Abstract}}
Two results are presented for reduced Yang-Mills integrals  with different symmetry groups and dimensions: the first is a
compact integral representation  in terms of the relevant variables
of the integral, the second is a method to analytically evaluate the integrals in cases
of low order. This is exhibited by evaluating a Yang-Mills integral over real symmetric
matrices of order $3$.

\section{Introduction}
In recent years the complete reduction of the $D$-dimensional $SU(N)$
Yang-Mills theory was studied by many authors. The reduced
supersymmetric partition function is an ordinary multiple integral
of the form
\begin{eqnarray}
Z=\int \prod_{\mu=1}^D  dX_{\mu} \; d\Psi \;e^{-S(X, \Psi)}
\label{a.1}
\end{eqnarray}
where the Euclidean action is
\begin{eqnarray}
S(X, \Psi)= -{\rm Tr}\, [X_\mu, X_\nu][X_\mu, X_\nu] -{\rm Tr}\, \Psi \, 
\Gamma^{\mu} \,[ X_{\mu}\, ,\,\Psi]
\label{a.2}
\end{eqnarray}
and the matrix-valued gauge potentials $X_{\mu}$ and their fermionic 
superpartners $\Psi$ have values in the $SU(N)$ Lie algebra
\begin{eqnarray}
&&X_\mu=X_\mu^a T^a \quad , \quad \Psi=\Psi^a T^a \quad \nonumber \\
&& [T^a, T^b]=i f^{abc}T_c \quad , \quad {\rm Tr} \,T^aT^b=\delta_{a b}
 \label{a.3}
\end{eqnarray}
There are several motivations to investigate these integrals both
in the maximally supersymmetric models, as well as in cases with
less supersymmetry \cite{IKKT} \cite{Windex} \cite{multi_ist} 
\cite{Staudacher:2000gx}. 
We refer to some recent papers \cite{Krauth:2000rc} 
\cite{Wheater-Austing} for introduction 
to the subject, review of recent results and further references. 
The interest in understanding these
 integrals is such that the study of simpler models without
 fermionic partners and for a variety of groups has been
considered very useful, 
also by means of extensive numerical investigations 
\cite{smallN-1} \cite{smallN-2} \cite{largeN} \cite{others} \cite{Sugino}.
 Unfortunately, a ge\-ne\-ral rigorous method for computing eq.~(\ref{a.1}) 
 is still unknown.  The development of techniques  to
understand these integrals for various values of $N$ and $D$ is
very desirable, as well as techniques which allow the analytic
evaluation of the integrals in cases of small order of the matrices. \\
In this letter we concentrate on the simpler, pure bosonic integrals of the form
\begin{eqnarray}
Z=\int \prod_{\mu=1}^D dX_{\mu}  \;e^{{\rm Tr}
 [X_\mu, X_\nu][X_\mu, X_\nu]}
\label{a.4}
\end{eqnarray}
The existence conditions and the convergence properties of~(\ref{a.1}) 
and (\ref{a.4}) (and of the corresponding correlation functions) 
have been analytically established in the most 
general case and for any compact semi-simple gauge group recently in 
\cite{Wheater-Austing}. As it is well known, these models,  
besides the global unitary invariance for similarity transformation of the matrices, have a rotational $SO(D)$ vector 
invariance. The latter
is the crucial ingredient for our
 two results : the first is a compact representation of eq.~(\ref{a.4})
in terms of its natural variables. It holds for generic values of $D$ and any
group. The second result is a technique to analytically evaluate the integrals
 in the case of small-order matrices. We exploit it in the evaluation of a $5D$ integral
corresponding to the Y-M integral over $3 \times 3$ real symmetric matrices.
It seems likely that both results should be helpful also for the supersymmetric
integrals.\\

It is useful to parametrize the random matrices with their independent entries.
 For instance we shall
consider the ensemble of real traceless symmetric matrices of order $3$ :
\begin{eqnarray}
 X^\mu =
\pmatrix { a^\mu & u^\mu  & v^\mu \cr
         u^\mu & -a^\mu -b^\mu  &  z^\mu  \cr
             v^\mu  & z^\mu  & b^\mu  \cr }  ; \quad \mu=1\ldots D
\label{a.5}
\end{eqnarray}
The action $S(X)$ is easily evaluated as a polynomial in the scalar products of the five
independent $D$-dimensional vectors.  The ensemble has the unitary invariance $SO(3)$ and
we evaluate, for $D>6 $
\begin{eqnarray}
 Z_{SO(3)}&=&\int d^D a \, d^D b\, d^D u \, d^D v\, d^D z \; e^{-S(X_{\mu})} = \nonumber\\
&=&\frac{\pi^{\frac{5D}{2}+1}\,3^{1-D}\,2^{6-\frac{7D}{2}} }{(D-4)(D-6) \,\Gamma
\left(\frac{3D}{4}-\frac{1}{2}\right) \Gamma \left(\frac{D}{2}-\frac{1}{2} \right)}
\label{a.6}
\end{eqnarray}
This should not be confused with the ensemble of real anti-symmetric 
matrices, also having
 the same $SO(3)$ unitary invariance. The latter ensemble is parametrized
 by three
$D$-dimensional vectors and corresponds to the only known integral which 
was exactly
evaluated in every dimension both in the bosonic and supersymmetric case
\cite{Windex}  \cite{Krauth:2000rc} \cite{Suyama}. 
Further evaluations were given for supersymmetric
integrals in fixed dimension by the deformation technique \cite{mns},
 reproducing the conjecture of \cite{green} and 
 afterwards numerically confirmed for small gauge groups in \cite{smallN-1}.

\section{A compact natural representation}
Let us consider an ensemble of matrices $X_{\mu}$, $\mu=1,\ldots ,D$,  parametrized by
   a number $n$ of $D$-dimensional real vectors.
For instance, in the familiar ensemble of Hermitian traceless
matrices of order $N$, invariant under the global $SU(N)$ group,
there are $n=N^2-1$ real vectors. The action $S=-{\rm
Tr}\,[X_{\mu},X_{\nu}]^2$ is a homogeneous polynomial in the
components of the vectors, and it only depends  on the $n(n+1)/2$
scalar products of the $n$ vectors, as it follows from the vector
$SO(D)$ invariance of the model. The bosonic partition function is
a multiple integral over $n D$ variables, with an integrand which
is a function of $n(n+1)/2$ variables. If $n \leq D$ the vectors
are generically linearly independent and it seems desirable to
express the partition function as an integral over the natural
variables, that is the scalar products. This is expressed by the
equation
\begin{eqnarray}
Z&=&\int \prod_{j=1}^n d {\vec v}_j\; F(
B)=\int_R \prod_{i \leq j} db_{ij}\;F(
B) \;I_{n,D}(B) \quad ,\nonumber\\
b_{ij} &\equiv& {\vec v}_i\cdot {\vec v}_j \equiv (v_i|v_j)=l_il_j \,z_{ij}
 \quad , \quad l_i \equiv\sqrt{ ({\vec v}_i)^2}
\label{b.1}
\end{eqnarray}
Here $F(B)$ stands for the usual Boltzmann weight (possibly multiplied by
a Pfaffian or a determinant resulting from the integration over the fermionic fields), the region of integration $R$ is described later in this section and
 $ I_{n,D}(B)$ is the joint probability density for a set
of $n$  vectors ${\vec v_i}$, with
given scalar products $ b_{ij}={\vec v_i}\cdot{\vec v_j}$
in the $D$-dimensional Euclidean space.\\
\underbar {Proposition 1}:
\begin{eqnarray}
 I_{n,D}(B)\!\!&\equiv & \!\!\int \prod_{i=1}^n d{\vec v_i}\;\prod_{1 \leq j < k \leq n}
\delta \left( b_{jk}-{\vec v_j}\cdot{\vec v_k}\right)
  = \nonumber\\ &=&
[{\det}_n(B)]^{(D-n-1)/2}\,\left( \prod_{j=1}^n   \frac{\Omega_{D-j}}{2}\right)
\; ,  \nonumber \\ {\rm where}\quad
  \Omega_{D-1}\!\!&=&\!\! 2 \frac{\pi^{D/2}}{\Gamma(D/2)}\quad , \quad
{\det}_n(B)
=(l_1l_2\ldots l_n)^2
 {\det}_n(Z)
\label{b.2}
\end{eqnarray}
 Proposition 1 may be proved from$\,$\footnote {A different proof of eq.~(\ref{b.2}) may also be given
by writing the integral representation for the delta functions and evaluating the integral
of a random real symmetric matrix in external field.}:\\
\underbar {Proposition 2}: for a given set of $n-1$ linearly
independent vectors $\{ {\vec v}_k\}$, $k=1,..,n-1$, in a real $D$-dimensional space,
$n \leq D$,
\begin{eqnarray}
i_n\!&=&\! \int d{\vec v}_n\, \delta\left(1-\frac{{\vec v}_n^2}{l_n^2}\right)\, \delta\left(z_{1n}-\frac{{\vec v_1}
\cdot{\vec v_n}}{l_1 l_n}\right)\ldots \delta \left(z_{n-1,n}- \frac{{\vec v_{n-1}}
\cdot{\vec v_n}}{l_{n-1}l_n}\right)=\nonumber\\
\!\!&=& \! \frac{\Omega_{D-n}}{2}\, l_n^D\, \frac{{\det}_n(Z)
^{(D-n-1)/2}}{{\det}_{n-1}
(Z)^{(D-n)/2}}=\frac{\Omega_{D-n}}{2} \,l_n^{n+1}\left( \prod_{j=1}^{n-1}l_j \right)
\; \frac{{\det}_n(B)^{(D-n-1)/2}}{{\det}_{n-1}
(B)^{(D-n)/2}}\nonumber\\
\label{b.3}
\end{eqnarray}
where $\det_{n-1}(B)$ is the leading principal minor obtained by deleting the last row and last column of $B$.\\
We outline here a proof for Proposition 2.
The vector ${\vec v}_n$ may be decomposed on the basis of the $n-1$ external vectors and
a $D-(n-1)$-dimensional vector ${\vec v}_{n \bot}$ in the orthogonal subspace :
 \begin{eqnarray}
{\vec v}_n&=&\omega_1 {\vec v}_1+..+\omega_{n-1} {\vec v}_{n-1}+{\vec v}_{n \bot} \quad ,
\nonumber \\   \int d{\vec v}_n&=& \sqrt{{\det}_{n-1}(B)} \,
\int d^{D-n+1} v_{n \bot}
\left(\prod_{j=1}^{n-1} \int_{-\infty}^{\infty}
 d\omega_j \right)
\label{b.4}
\end{eqnarray}
The vector ${\vec v}_{n \bot}$ only occurs in the first delta function
$ \delta(1-  {\vec v}_n^2/l_n^2)
$, and its integral is trivial
 \begin{eqnarray}
\!\!\!\lefteqn{\!\!\!\int d^{D-n+1} {\vec v}_{n \bot}\,\delta \left(1-\frac{ {\vec v}_{n \bot}^2+
(\sum \omega_j {\vec v}_j)^2}{l_n^2}\right) = \Omega_{D-n} \int_0^{\infty} d\rho\, \rho^{D-n}
\delta \left(1-\frac{ \rho^2+
(\sum \omega_j {\vec v}_j)^2}{l_n^2}\right)=} \nonumber \\
&& =\frac{\Omega_{D-n}}{2}\,l_n^2\; \left[ l_n^2- \left( 
\sum \omega_j {\vec v}_j \right)^2\right]^{\frac{D-n-1}{2}} \, ; \quad  
\left(\sum_{j=1}^{n-1}\omega_j {\vec v}_j \right)^2= \sum_{j,k=1}^{n-1} \omega_j \omega_k \,b_{jk} \nonumber \\
  \label{b.5}
\end{eqnarray}
The remaining $n-1$ delta functions fix the $\omega_j$ as the solution of the system
 $
\sum_{j=1}^{n-1}b_{ij}\, \omega_j=l_il_n\,z_{in}=b_{in}$, $i=1,..,n-1$, that is
${\bar \omega}_j=\sum_s [B^{-1}]_{j s}b_{sn}$. Then
 \begin{eqnarray}
&&l_n^2- \left(\sum \omega_j {\vec v}_j \right)^2=b_{nn}-\sum_{j,k=1}^{n-1}
{\bar \omega}_j {\bar \omega}_k \,b_{jk}=b_{nn}-\sum_{j=1}^{n-1}
{\bar \omega}_j \,b_{jn}=\nonumber\\ &&=
b_{nn}-\sum_{j,s=1}^{n-1}b_{ns}[B^{-1}]_{sj} \,b_{jn}=\frac{{\det}_n(B)}
{{\det}_{n-1}(B)}
  \label{b.6}
\end{eqnarray}
Finally Proposition 2 is proved after using
 \begin{eqnarray}
\int_{-\infty}^{\infty}\prod_{j=1}^{n-1}d\omega_j\,\prod_{k=1}^{n-1}\delta\left(
z_{kn}-\frac{\sum_{r=1}^{n-1}\omega_r\,l_r\,z_{kr}}{l_n}\right)=\frac{l_n^{n-1}}
{l_1l_2..l_{n-1}}\frac{1}{{\det}_{n-1}(Z)}
  \label{b.7}
\end{eqnarray}
Proposition 1 follows now from Proposition 2
since the joint density $I_{n,D}(B)$, from its definition, may be written as the
product of the integrals $i_k$.

The cosines $z_{ij}$ are bound by $-1 \leq z_{ij} \leq 1$ but the derivation of
eq.~(\ref{b.2}) implies the stricter bound that $\det_k(Z)>0$ for each $k$, $k=1,..,n$, which is necessary and sufficient condition for the real symmetric matrix $Z$ to have all the eigenvalues positive. \\
Eqs.(\ref{b.1}), (\ref{b.2}) provide a very simple reduction of the partition function
to the smaller number of relevant variables which has general validity for
 different matrix ensembles and dimensions. It may be useful for numerical integration
and possibly  for more general purposes when averaging over random vectors.\footnote
 {See for instance chapter 21 of ref.\cite{meh}, where one studies averages over random
 orthogonal vectors. Our eq.~(\ref{b.2}) for $n=2$, $z_{12}=0$ , reproduces the result
eq.~(21.1.11) of   \cite{meh}.} The partition function expressed as
multiple integral over the scalar products is closely related to
the Ingham-Siegel integral \cite{fyod}. In fact the well known
Yang-Mills integral for the $SU(2)$ invariant ensemble is a
trivial case. After parametrization of the hermitian traceless
matrices of order $2$ with $3$ real entries, one obtains
 \begin{eqnarray}
Z_{SU(2)}\!\!&=&\!\! \int \prod_{j=1}^3 d {\vec v}_j \;e^{-S_{SU(2)}}=
\nonumber\\ &=&
\frac{\pi^{3(D-1)/2}}{
\Gamma\left(\frac{D}{2}\right)\Gamma\left(\frac{D-1}{2}\right)\Gamma\left(\frac{D-2}{2}
\right)} \int_{B>0}  dB\;e^{-S_{SU(2)}}
 \;[\det(B)]^{(D-4)/2} \quad ,
 \nonumber 
\end{eqnarray}
where
\[
S_{SU(2)}=16 \bigg(b_{11}b_{22}+b_{11}b_{33}+
b_{22}b_{33}-b_{12}^2-b_{13}^2-b_{23}^2\bigg)=
16 \det(B)\;{\rm tr}\,(B^{-1}) \, \quad  .
\]
We change integration matrix variables $C_{ij}=(\det B)[B^{-1}]_{ij}$,
 $dB=(\det C)^{-1}dC$, $\det B=(\det C)^{1/2}$
  \begin{eqnarray}
 \!&&\!\!\int_{B>0}  dB\;e^{-S_{SU(2)}}
 \;[\det(B)]^{(D-4)/2} =
2^{-3D}\int_{C>0} dC\;(\det C)^{(D-8)/4}\;e^{-{\rm tr}\, C}=\nonumber\\
&&=\frac{\pi^{3/2}}{2^{3D}} \Gamma\left(\frac{D}{4}\right)
 \Gamma\left(\frac{D-1}{4}\right)
 \Gamma\left(\frac{D-2}{4}\right)
  \label{b.8}
\end{eqnarray}
Since for the $SU(2)$ case in $D=10$ the Pfaffian arising from the fermion integration
is proportional to $(\det B)^4$, also the supersymmetric $SU(2)$ integral is easily
obtained, reproducing the result of \cite{Windex} \cite{Suyama}.
 \vskip 1 cm

 \section{Reduction by iterated projections}

 A useful technique which avoids the inconvenience of the Jacobian and
positivity requirements is the {\sl {Reduction by iterated projections}}.
The price is a more complicated expression of the action. To put it on
general ground, let us consider the integral in $n$ real vector variables
of a function which only depends on the scalar products in $D$ dimensions
\begin{eqnarray}
 I= \int d^Dx_1 d^Dx_2 \ldots d^Dx_n F\bigg( (x_i|x_j)_D;  i\le j \bigg)
  \label{c.1}
\end{eqnarray}
\underbar {Proposition 3}\parno
Case $D\ge n$:
\begin{eqnarray}
 I=\Omega_{D-1}\ldots \Omega_{D-n}\int_0^\infty dx_1 x_1^{D-1} \ldots
\int_0^\infty dx_n x_n^{D-n} \int_{-\infty}^\infty \prod_{i<j} dx_{ij}
\, F(\ldots )
\end{eqnarray}

\begin{eqnarray}
 (x_i|x_j)_D = x_{1i}x_{1j} + \ldots + x_{i-1,i}x_{i-1,j}+x_ix_{ij} ,\quad
i\le j
\end{eqnarray}
\noindent
Case $D<n$:
\begin{eqnarray}
 I=\Omega_{D-1}\ldots \Omega_0 \int_0^\infty dx_1 x_1^{D-1} \ldots
\int_0^\infty dx_D  \int_{-\infty}^\infty \prod_{ {i<j}\atop {i\le D}}
dx_{ij}\,F(\ldots )
\end{eqnarray}

\begin{eqnarray}
&& (x_i|x_j)_D = x_{1i}x_{1j} + \ldots + x_{Di}x_{Dj} 
,\quad D<i\le j \quad ;\nonumber\\
 &&(x_i|x_j)_D = x_{1i}x_{1j} + \ldots + x_{i}x_{ij} ,\quad i<j, \, i\le D
\end{eqnarray}

Proof: The rotational invariance in the space $R^D$ of all vectors allows
us to choose the vector $x_1$ as the polar direction. Any other vector $x_i$
is then described by its component parallel to $x_1$, denoted as $x_{1i}$
living in $R$, and a vector orthogonal to it, which (with abuse of language)
we again denote $x_i$,  living in the space $R^{D-1}$.
Thus we have the first sweep
\begin{eqnarray}
 I= \Omega_{D-1} \int_0^\infty dx_1 x_1^{D-1}
\int_{-\infty}^\infty dx_{12}\ldots dx_{1n}
\, \int dx_2^{D-1}\ldots dx_n^{D-1} F (\ldots )
\end{eqnarray}
\noindent
where the scalar products in $F$ are easily evaluated:
\begin{eqnarray}
 (x_1|x_1)_D=x_1^2, \quad   (x_1|x_i)_D=x_1 x_{1i}, \quad
(x_i|x_j)_D=x_{1i}x_{1j}+(x_i|x_j)_{D-1},   \quad 1<i\leq j
\nonumber
\end{eqnarray}
Next we choose the vector $x_2$ as the polar direction in $R^{D-1}$,
and introduce accordingly new real variables $x_{2i}$ and vectors
$x_i$ in $R^{D-2}$, $i=3,\ldots n$. The integral is now:
\begin{eqnarray}
 I&=&\Omega_{D-1}\Omega_{D-2} \int_0^\infty dx_1 x_1^{D-1}
\int_0^\infty  dx_2 x_2^{D-2} \int_{-\infty}^{\infty} dx_{12}\ldots dx_{1n}
dx_{23}\ldots dx_{2n} \nonumber\\ &&\quad
\int d^{D-2}x_3 \ldots d^{D-2}x_n \;F(\ldots )
\end{eqnarray}
 The scalar products in $R^{D-1}$ that still appear in $F$ are evaluated:
\begin{eqnarray}
 (x_2|x_2)_{D-1}=x_2^2, \quad  (x_2|x_i)_{D-1}= x_2x_{2i} ,\quad
(x_i|x_j)_{D-1}=x_{2i}x_{2j}+ (x_i|x_j)_{D-2}, \quad  2<i\le j
\nonumber
\end{eqnarray}
The process is iterated and we have two situations: $D\ge n$, $D<n$.
In the first case we can iterate the process $n$ times, to obtain an
integral in $n$ positive variables $x_1$ ... $x_n$ and ${1\over 2}n(n-1)$
real variables $x_{ij}$, $i<j$. When $D<n$ the process can be iterated only
$D$ times; we obtain $D$ positive variables $x_1$ ... $x_D$,
and ${1\over 2}D(2n-D-1)$ real variables $x_{ij}$, $i<j$, $i=1\ldots D$.

\section{Y-M integral with $3\times 3$ real symmetric matrices}

The $ 3\times 3$ traceless real symmetric matrices of the Yang-Mills integral are
parametrized as indicated in eq.~(\ref{a.5}). The partition function
\begin{eqnarray}
 Z_{SO(3)}=\int d^D a \, d^D b\, d^D u \, d^D v\, d^D z \; e^{4S_0}
\end{eqnarray}
contains the following expression for the action, in terms of scalar products like
$(a|b)_D=\sum_\mu a^\mu b^\mu $:
\begin{eqnarray}
 S_0 \!\!&=\!\!& -4 [a^2 u^2 -  (a|u)^2 ] -   [ a^2 z^2 -  (a|z)^2 ] -  [a^2 v^2 - (a|v)^2 ]
               - [b^2 u^2 -  (b|u)^2 ] \nonumber\\ &&
- 4 [ b^2 z^2 -  (b|z)^2 ] -  [b^2 v^2 - (b|v)^2 ]
               - [u^2 z^2 -  (u|z)^2 ] -  [ v^2 z^2 -  (v|z)^2 ] \nonumber \\
&& - [u^2 v^2 - (u|v)^2 ]
               -4 [u^2 (a|b) - (u|a)(u|b)]  -4 [z^2 (a|b) - (z|a)(z|b)]
                \nonumber \\
&& +2 [v^2 (a|b) - (v|a)(v|b)] 
               +6(a|z)(u|v) -6(a|v)(u|z) \nonumber \\
&& + 6(b|u)(v|z)-6(b|v)(u|z)
\end{eqnarray}
The huge invariance of the action can be exploited by reducing to a smaller number of effective variables. We have been able to compute
$Z_{SO(3)}$ by using the technique of
successive projections, described in the previous Section.\\
For $D\ge 5$ the reduction yields a 15-dimensional integral:
\begin{eqnarray}
\!\!Z_{SO(3)}\!\!\!\! &=&\!\!\!\! \Omega_{D-1}\ldots  \Omega_{D-5}
     \int_0^\infty dv  v^{D-1}  \int_0^\infty dz  z^{D-2} \int_0^\infty du  u^{D-3}
\int_0^\infty da  a^{D-4}      \int_0^\infty db  b^{D-5} \nonumber\\ &&
\int_{-\infty }^\infty dz_v\,  du_v\, da_v\, db_v\, du_z\, da_z\, db_z \, da_u\,
db_u\, db_a \; e^{S_1+ S_2 +S_3+S_4}
\label{r.1}
\end{eqnarray}
where the action in the reduced variables has been decomposed in terms
with the following meaning. $S_1$ includes all terms in the
variables $a$, $b_a$ and $b$, that terminated  the reduction procedure
and can be integrated immediately:
\begin{eqnarray}
 S_1 &=&   -4a^2 [v^2 + (z^2 +z_v^2) +4 (u^2+ u_v^2 +u_z^2)] \nonumber \\
     &&  -4(b^2+b_a^2) [v^2+ 4(z^2 +z_v^2) +(u^2+ u_v^2 +u_z^2)] \nonumber \\
    &&  +8a b_a [v^2 -2 (z^2 +z_v^2 +u^2+ u_v^2 +u_z^2)]
\end{eqnarray}

The integration yields a factor that does not depend on the 6 variables
$a_x$, $b_x$, where $x=u,v,z$. These variables appear in the action
quadratically  and linearly. The quadratic terms are collected in $S_2$ and
the linear terms in $S_3$, in a form suitable for a Gaussian integration:
 \begin{eqnarray}
 S_2 = - 4 X^t M X \quad ; \quad S_3 = 8vX^tY \end{eqnarray}
  $$ {\rm where} \qquad
 X=\pmatrix { a_v+2b_v \cr 2a_v + b_v \cr
                           a_z+2b_z  \cr 2a_z + b_z \cr
                           a_u+2b_u \cr 2a_u+ b_u\cr}\quad , \quad
    Y =\pmatrix {-zu_z     \cr -zu_z\cr
                 2u_zz_v -zu_v\cr -u_zz_v+ 2zu_v\cr
                           2uz_v\cr -uz_v\cr } \quad , $$
$$  M=\pmatrix {
 z^2    &  0 & -zz_v & 0  & 0 & 0 \cr
  0 & u^2+u_z^2 & 0 & -u_z u_v & 0  &-uu_v \cr
  -zz_v & 0 &  v^2+z_v^2 & -v^2 & 0  & 0    \cr
   0   &-u_z u_v   & -v^2 & v^2+u^2+u_v^2 & 0 & -uu_z \cr
   0  & 0 &   0&  0  & v^2+z^2+z_v^2  & -v^2 \cr
   0  &-uu_v &   0 & -uu_z & -v^2 & v^2 + u_v^2 + u_z^2 \cr }$$
The remaining part $S_4$ of the action contains terms in the reduced
variables originating from the off diagonal matrix elements, namely
$v$, $z$, $z_v$, $u$, $u_v$ and $u_z$:
\begin{eqnarray}
 S_4= -4z^2(v^2+u_v^2)-4u_z^2(v^2+z_v^2)-4u^2(v^2+z^2+z_v^2)
+8zu_zu_vz_v  \nonumber\\
\end{eqnarray}
As we described, the $Z_{SO(3)}$ integral can be done in independent steps. We have a
factor coming from the  triple integral:
\begin{eqnarray}
 I_1\!& =&\!\int_0^\infty da \, a^{D-4}\int_0^\infty b^{D-5}db
\int_{-\infty }^\infty db_a \;e^{S_1}= \nonumber\\
 &=&\! {{\sqrt \pi}\over 4}(12)^{-D+3}\frac{
      \Gamma \left ( {{D-3}\over 2}\right)  \Gamma \left ( {{D-4}\over 2}\right )}{
 [v^2(u^2+u_v^2+u_z^2+z^2+z_v^2)+(z^2+z_v^2)(u^2+u_v^2+u_z^2)]^{(D-3)/2} }
\nonumber\\
\end{eqnarray}
\noindent
At the same time one can do the Gaussian integral
 \begin{eqnarray}
I_2 \!&=&\!\int_{-\infty}^{\infty } da_v db_v da_z db_z da_u
db_u \;e^{S_2+S_3}= \nonumber\\ &=&  {1\over {12^3}} \int d^6 X \,e^{-X^tMX
+4vX^tY} = {{\pi^3}\over {12^3}} (\det M)^{-1/2} \, e^{S_5}  \end{eqnarray}
The computations for inverting the matrix $M$ and evaluating the
quadratic form $S_5= 4v^2 Y^t M^{-1}Y$ were done with the aid of
Mathematica:
$$  S_5  = 4(v^2+z_v^2)(u^2+u_z^2) + 4z^2(v^2+u_v^2)- 8zz_vu_vu_z
- 4{{(vzu)^2}\over {z^2+z_v^2}} - 4 {{(vzu)^2}\over {u^2+u_v^2+u_z^2}} \quad ,
$$
$$ \det M = v^4 z^2 u^2 (z^2+z_v^2)(u^2+u_v^2+u_z^2) $$
In summing the expressions $S_4$ and $S_5$, to obtain the residual
action, many terms cancel:
 \begin{eqnarray}
S_4+S_5 = -4 z^2 u^2 - 4v^2z^2u^2 \frac
{u^2+u_v^2+u_z^2+z^2+z_v^2} {(z^2+z_v^2)(u^2+u_v^2+u_z^2)} \end{eqnarray}
We arrive at the following stage of computation of the full
integral:
  \begin{eqnarray}
Z_{SO(3)} ={1\over 4} \pi^{7/2} 12^{-D} \Gamma\left({D\over
2}-2\right)\Gamma\left({D\over 2}-{3\over 2}\right)\, \Omega_{D-1}\ldots \Omega_{D-5}
\;I_3 \end{eqnarray}
where we still have to  evaluate
\begin{eqnarray}
 I_3 \!&=& \!
\int_0^\infty dv v^{D-3} \int_0^\infty dz z^{D-3} \int_0^\infty du
u^{D-4} \int_{-\infty}^\infty dz_v \, du_v \, du_z \, e^{S_4+S_5}
\nonumber\\ &&
[v^2(u^2+u_v^2+u_z^2+z^2+z_v^2)+(z^2+z_v^2)(u^2+u_v^2+u_z^2)
]^{-{1\over 2}(D-3)}
\nonumber\\ &&
[(u^2+u_z^2+u_v^2)(z^2+z_v^2)]^{-1/2}\nonumber
\end{eqnarray}
Since the variables $u_v$ and $u_z$ always appear in the
combination $u_v^2+u_z^2 =\rho^2 $, we introduce polar
variables obtaining a factor $2\pi$ from angle integration. It is
also convenient to change from the variable $v$ to the variable
$x$:
 \begin{eqnarray} x = v \sqrt { \frac{
   u^2+\rho^2 +z^2+z_v^2 } {(u^2+\rho^2 )(z^2+z_v^2)}}  \nonumber
   \end{eqnarray}
\begin{eqnarray} I_3 \!&=& \!2\pi \int_0^\infty dx \, x^{D-3} (1+x^2)^{-
\frac{D-3}{2}}
 \int_0^\infty dz \, z^{D-3} \int_0^\infty du \,u^{D-4} e^{-4u^2z^2(1+x^2)}
 \nonumber\\ && \quad
\int_0^\infty \rho \, d\rho
 \int_{-\infty}^\infty dz_v (
 u^2+\rho^2+z^2+z_v^2)^{-{D\over 2} +1}
 \end{eqnarray}
The double integral in $\rho $ and $z_v$ is elementary, and poses the
restriction $D>5$ for convergence,
\begin{eqnarray}
\int_{-\infty}^{\infty} \!dz_v \int_0^\infty \! \rho\, d\rho \,(u^2+\rho^2
+z^2+z_v^2)^{-{D\over 2} +1} =(z^2+u^2)^{-{D\over 2}+{5\over 2}}
B\left( \frac{D-5}{2}, {3\over 2} \right) \nonumber 
\end{eqnarray}
The remaining integrals are easily done with the substitution $u=
(1+x^2)^{-1/4} r\cos\theta $ and $z=(1+x^2)^{-1/4} r\sin\theta $. We obtain,
with the further restriction $D>6$ for convergence,
\begin{eqnarray}
I_3\!\!&=&\!\! 2\pi B\left( \frac{D-5}{2}, {3\over 2} \right) \int_0^\infty \!
dx\, x^{D-3} (1+x^2)^{-{3\over 4}D+{3\over 2}} \int_0^{\pi/2} \! d\theta
\cos^{D-4}\theta \sin^{D-3}\theta  \nonumber\\ && \qquad
\int_0^\infty \! dr \, r^{D-1}
e^{-4r^4\sin^2\theta\cos^2\theta }
=4 \pi^{5/2} 2^{-{3\over2}D}  {{\Gamma ({D\over 2}-1)
\Gamma ({D\over 2}-3) }\over { \Gamma ({3\over 4}D- {3\over 2})}}\nonumber\\
\end{eqnarray}
The final result , for $D>6$ is:
\begin{eqnarray}
Z_{SO(3)} = \frac{\pi^{\frac{5D}{2}+1}\,3^{1-D}\,2^{6-\frac{7D}{2}} }{(D-4)(D-6) \,\Gamma
\left(\frac{3D}{4}-\frac{1}{2}\right) \Gamma \left(\frac{D}{2}-\frac{1}{2} \right)}
\end{eqnarray}
In the case $D=4$, one should truncate the reduction by projections, as described in the
previous section (since the $5$ vectors are not linearly independent).   This produces a set of $14$ integration variables and an action which, in simple terms, are obtained  from
eq.~(\ref{r.1}) by suppressing the integral in $b$ and all
terms containing $b$ in $S_1$. We checked that also
 in this case  the integrations lead to a divergent result.

\vskip 1 cm

\subsection*{Acknowledgments}
One author (G.M.C.) wishes to thank M.Bonini, R.De Pietri, M.P.Manara for useful discussions, another author (G.V.) wishes to thank J.Wheater for sti\-mu\-la\-ting discussions and comments. The work of G.V. is supported by the EU network on ``Discrete Random Geometry'', grant HPRN-CT-1999-00161.

\end{document}